\documentclass[preprintnumbers,article,amsmath,amssymb,floatfix,10pt,prd,twocolumn,superscriptaddress,nofootinbib]{revtex4-2}
\usepackage{bm}
\usepackage{amsfonts}
\usepackage{latexsym}
\usepackage[latin1]{inputenc}
\usepackage{graphicx}
\usepackage{amsmath}
\usepackage{palatino}
\usepackage{mathpazo}
\usepackage{textcomp}
\usepackage{float}
\usepackage{booktabs}
\usepackage{dcolumn}
\usepackage{ragged2e}
\usepackage{hyperref}
\hypersetup{colorlinks,citecolor=blue}
\hypersetup{colorlinks=true,linkcolor=red,filecolor=magenta,    urlcolor=blue}
\usepackage{amsmath}
\usepackage{xcolor}
\usepackage{orcidlink}
\usepackage{epsfig}
\usepackage{caption}
\usepackage{subcaption}
\usepackage{commath}
\def\jnl@style{\it}
\def\aaref@jnl#1{{\jnl@style#1}}

\def\aaref@jnl#1{{\jnl@style#1}}

\def\aj{\aaref@jnl{AJ}}                   
\def\apj{\aaref@jnl{ApJ}}                 
\def\apjl{\aaref@jnl{ApJ}}                
\def\apjs{\aaref@jnl{ApJS}}               
\def\apss{\aaref@jnl{Ap\&SS}}             
\def\aap{\aaref@jnl{A\&A}}                
\def\aapr{\aaref@jnl{A\&A~Rev.}}          
\def\aaps{\aaref@jnl{A\&AS}}              
\def\mnras{\aaref@jnl{Mon.~Not.~Roy.~Astron.~Soc.}}             
\def\prd{\aaref@jnl{Phys.~Rev.~D}}        
\def\prc{\aaref@jnl{Phys.~Rev.~C}}  
\def\prl{\aaref@jnl{Phys.~Rev.~Lett.}}    
\def\qjras{\aaref@jnl{QJRAS}}             
\def\skytel{\aaref@jnl{S\&T}}             
\def\ssr{\aaref@jnl{Space~Sci.~Rev.}}     
\def\zap{\aaref@jnl{ZAp}}                 
\def\nat{\aaref@jnl{Nature}}              
\def\aplett{\aaref@jnl{Astrophys.~Lett.}} 
\def\apspr{\aaref@jnl{Astrophys.~Space~Phys.~Res.}} 
\def\physrep{\aaref@jnl{Phys.~Rep.}}      
\def\physscr{\aaref@jnl{Phys.~Scr}}       
\def\commat{\aaref@jnl{Comm.~Math.~Phys.}}              
\def\science{\aaref@jnl{Science}}               
\def\cqg{\aaref@jnl{Classical Quant.~Grav.}}            
\def\jpcs{\aaref@jnl{JPCS}}                                     
\def\ijmpd{\aaref@jnl{Int.~J.~Mod.~Phys.~D}}                    
\def\grg{\aaref@jnl{Gen.~Relat.~Gravit.}}               
\def\rpp{\aaref@jnl{Rep.~Prog.~Phys.}}          
\def\npa{\aaref@jnl{Nucl.~Phys.~A}}        
\def\lrr{\aaref@jnl{Living Rev.~Rel.}}                   
\def\jcap{\aaref@jnl{J.~Cosmology Astropart.~Phys.}}    
\def\rmp{\aaref@jnl{Rev.~Mod.~Phys.}}   
\def\epjc{\aaref@jnl{Eur.~Phys.~J.~C}} 
\def\plb{\aaref@jnl{~Phy.~Lett.~B}} 
\def\mpla{\aaref@jnl{Mod.~Phy.~Lett.~A}} 
\def\arxiv{\aaref@jnl{arxiv.org}}

\allowdisplaybreaks[1]

\begin{document}
\color{black}       
%
\title{Reconstruction of the singularity-free $f(\mathcal{R})$ gravity via Raychaudhuri equations}

\author{Gaurav N. Gadbail\orcidlink{0000-0003-0684-9702}}
\email{gauravgadbail6@gmail.com}
\affiliation{Department of Mathematics, Birla Institute of Technology and
Science-Pilani,\\ Hyderabad Campus, Hyderabad-500078, India.}

\author{Simran Arora\orcidlink{0000-0003-0326-8945}}
\email{dawrasimran27@gmail.com}
\affiliation{Department of Mathematics, CDOE, Chandigarh University, Punjab, 140413, India.}

\author{P.K. Sahoo\orcidlink{0000-0003-2130-8832}}
\email{pksahoo@hyderabad.bits-pilani.ac.in}
\affiliation{Department of Mathematics, Birla Institute of Technology and
Science-Pilani,\\ Hyderabad Campus, Hyderabad-500078, India.}

\author{Kazuharu Bamba\orcidlink{0000-0001-9720-8817}}
\email{bamba@sss.fukushima-u.ac.jp}
\affiliation {Faculty of Symbiotic Systems Science,
Fukushima University, Fukushima 960-1296, Japan}
%

\begin{abstract}
We study the bounce cosmology to construct a singularity-free $f(\mathcal{R})$ model using the reconstruction technique. The formulation of the $f(\mathcal{R})$ model is based on the Raychaudhari equation, a key element employed in reconstructed models to eliminate singularities. We explore the feasibility of obtaining stable gravitational Lagrangians, adhering to the conditions $f_{\mathcal{R}}>0$ and $f_{\mathcal{R}\mathcal{R}}>0$. Consequently, both models demonstrate stability, effectively avoiding the Dolgov-Kawasaki instability. Our assessment extends to testing the reconstructed model using energy conditions and the effective equation-of-state (EoS). Our findings indicate that the reconstructed super-bounce model facilitates the examination of a singularity-free accelerating universe for both phantom and non-phantom phases. However, in the case of the reconstructed oscillatory bounce model, two scenarios are considered with $\omega=-1/3$ and $\omega=-2/3$. While the model proves suitable for studying a singular-free accelerating universe in the $\omega=-1/3$ case, it fails to demonstrate such behavior under energy conditions for the $\omega=-2/3$ scenario. The reconstructed models accommodate early-time bouncing behavior and late-time cosmic acceleration within a unified framework.

\end{abstract}
\maketitle
\date{\today}

\section{Introduction}
Understanding cosmic acceleration is crucial for gaining insights into the nature of the universe. Several proposed scenarios aim to clarify this remarkable discovery, falling into two main categories. In one approach, introducing an exotic component known as dark energy is suggested to be the driving force behind cosmic acceleration. Up to the present, every independent cosmological observation agrees with the existence of either vacuum energy density or a non-zero cosmological constant $\Lambda$ \cite{Riess}. However, the straightforward cosmological constant model still faces challenges, particularly the coincidence problem and the fine-tuning problem \cite{Abbott}. These issues have prompted extensive efforts to grasp the early stages of the evolution of the universe. On the other side, achieving accelerated expansion is possible by modifying the gravitational aspect of the Einstein-Hilbert action \cite{Carroll}. In line with this approach, numerous extended theories of gravity have been suggested in recent decades, including a nonminimally coupled scalar field theory \cite{Bertolami}, scalar-tensor theory \cite{Alonso}, and $f(\mathcal{R})$ gravity \cite{f(R)}. Particularly, $f(\mathcal{R})$ gravity theories, where $f(\mathcal{R})$ is a function of the Ricci scalar $\mathcal{R}$, are widely utilized in contemporary cosmology. Higher powers of $\mathcal{R}$ in these models can explain early-stage accelerated expansion known as inflation, while exploring negative powers of $\mathcal{R}$ may offer insights into late-time acceleration in the universe \cite{Staro/1980}. To validate these theoretical models, rigorous testing against observational data and various criteria, including solution stability and local astronomical phenomena, is essential \cite{Dolgov}. For a comprehensive review of $f(\mathcal{R})$ gravity models, covering their compatibility with observations and theoretical considerations, check the following references \cite{Sotiriou}. In summary, $f(\mathcal{R})$ gravity models present an avenue for alternative gravity theories, necessitating thorough examination against observational and theoretical benchmarks for validation. They are appreciated for their capacity to portray the distribution of large-scale structures in the universe and provide explanations for the observed cosmic acceleration during later cosmic epochs \cite{Song} and the solar system predictions \cite{Solar}.\\
The intricate nature of the fourth-order field equations complicates efforts to comprehend the physics of these theories, making it challenging to obtain both exact and numerical solutions for comparison with observations. Nonetheless, recent advancements have tackled this complexity using various effective techniques. A notable approach, grounded in the theory of dynamical systems, has shown considerable success \cite{Odintsov}. This method offers a simplified way to obtain exact solutions and provides a qualitative overview of the overall dynamics of these models. Another interesting approach is the reconstruction method. In this method, one starts with the assumption that the expansion history of the universe is precisely understood. The field equations are then inverted to identify the particular category of $f(\mathcal{R})$ theories associated with a given FLRW model. This method demonstrates numerous instances where the reconstruction of $f(\mathcal{R})$ gravity aligns with well-known cosmological phases. These instances encompass the $\Lambda$CDM epoch, epochs of deceleration/acceleration resembling the presence of phantom and non-phantom matter, late-time acceleration involving the crossing of the phantom-divide line, a transient phantom epoch, and scenarios depicting an oscillating universe \cite{Carloni}.

The Raychaudhuri equation is commonly seen as a highly refined equation that has significantly contributed to exploring Einstein's general theory of relativity. Raychaudhuri formulated his well-known equation, characterized by its pure geometric nature and absence of a direct association with the Einstein field equation \cite{Raychaudhuri}. The Raychaudhuri equation addresses the concept of geodesic congruence, aiming to explore and understand the temporal evolution of a congruence of geodesics as perceived by an adjacent member within the exact congruence. In essence, the equation elucidates the relationship between the rate of change in the scalar expansion of a test particle and the gravitational impact associated explicitly with the energy-momentum of that test particle. 

Considering the geometric perspective, we aim to create a mathematically versatile analytical solution for different epochs within the $f(\mathcal{R})$ gravity model. Since the Raychaudhuri equation is entirely based on geometry, we find it applicable to our case. We have opted for two types of cosmological bouncing scale factors: super bounce and oscillatory bounce-to establish the functional form of $f(\mathcal{R})$ gravity.
Here are some advantages associated with the super-bounce and oscillatory bouncing models: the avoidance of singularities is a key benefit; the cyclic nature of the model provides an alternative to the standard cosmological paradigm; compatibility with modified gravity theories contributes to the avoidance of singularities; the cyclic evolution may naturally address cosmological problems such as the horizon and flatness problems; and the super-bounce and oscillatory bounce model provides an alternative to inflationary cosmology. Utilizing the Raychaudhuri equation, we examine effective energy conditions and phases dominated by dark energy, aiming to formulate a general yet practical $f(\mathcal{R})$ gravity model as an analytical solution. It is important to note that we are aware of the prior derivation of the Raychaudhuri equation in the context of $f(\mathcal{R})$ gravity using a different approach.

In recent years, there has been a notable interest in cosmological bouncing solutions as an alternative to the conventional Big Bang singularity, aiming to address the unnaturalness associated with the origin of the universe \cite{Nojiri/2017,Haro/2023}. Nonsingular bouncing cosmologies, in particular, have gained attention for their potential to resolve the singularity problem linked to the prevalent inflationary scenario in early universe cosmology \cite{Mukhanov}. Achieving a bouncing solution for the spatially flat FLRW metric within general relativity (GR) typically involves introducing matter components that violate the null energy condition (NEC), such as ghost fields, ghost condensates, or Galileons \cite{Gal}. However, for those seeking to avoid exotic matter components and still realize a bouncing solution in the spatially flat FLRW metric, modified gravity emerges as a viable option \cite{BounceM}. It is reasonable to anticipate modifications to general relativity in the high-curvature regime near a curvature singularity. Various attempts have been made to realize a bouncing scenario within the $f(\mathcal{R})$ theories of gravity \cite{Chakraborty}. \\
In subsequent work, we use a reconstruction technique to analyze different cosmological bounce scenarios within the framework of $f(\mathcal{R})$ gravity. In particular, we develop $f(\mathcal{R})$ gravity models that exhibit late-time cosmic acceleration and a bounce in the early universe. We look into these models' stability. We investigate two kinds of bounces: super-bounce and oscillatory bounce. The reconstruction approach first uses the explicit form of the scale factor related to a specific bounce type, enabling the calculation of the Hubble parameter $H(t)$. Next, we directly apply the Raychaudhuri equation to determine the $f(\mathcal{R})$ gravity model for the two cases. In the first case, we get a combination of powers of the Ricci scalar $\mathcal{R}$, which represents an eternally accelerated model. Combinations of hypergeometric functions appear in the second scenario. We also assess the viability of the $f(\mathcal{R})$ models against various cosmological requirements.\\
The paper is structured as follows: In Section \ref{section II}, a concise overview of $f(\mathcal{R})$ gravity theory is presented. Moving to Section \ref{section III}, we derive the Raychaudhuri equation for $f(\mathcal{R})$ gravity within the framework of the FLRW metric. Section \ref{section IV} is dedicated to the reconstruction of suitable gravitational Lagrangians $f(\mathcal{R})$ that can accurately replicate the cosmic evolution observed in super-bounce and oscillatory bounce models. Additionally, we assess the validity criteria of these $f(\mathcal{R})$ Lagrangians. The cosmological aspects of the reconstructed models are examined in Section \ref{section v}. Finally, Section \ref{section VI} provides the conclusions drawn from the study.

\section{$f(\mathcal{R})$ gravity}
\label{section II}
We begin by recalling the straightforward modification of the Einstein-Hilbert action, where $\mathcal{R}$ is replaced with a general function of the Ricci scalar, namely \cite{f(R)}
\begin{equation}
\label{action}
S_{\mathcal{R}}=\int \sqrt{-g}\left(f(\mathcal{R})+\mathcal{L}_m\right) d^4x,
\end{equation}
where $f(\mathcal{R})$ is an analytic function of the Ricci scalar $\mathcal{R}$ and $\mathcal{L}_m$ is the matter Lagrangian. Here we take $8\pi\,G=1$. The governing field equations for a general $f(\mathcal{R})$ theory is derived by varying the action \eqref{action} with respect to the metric alone, yielding the following field equations
\begin{equation}
\label{fe}
    f_{\mathcal{R}}R_{\mu\nu}-\frac{f}{2}g_{\mu\nu}-\left(\nabla_{\mu}\nabla_{\nu}-g_{\mu\nu}\square\right)f_{\mathcal{R}}=T_{\mu\nu},
\end{equation}
where $f_{\mathcal{R}}=df/d\mathcal{R}$ and $T_{\mu\nu}\equiv-\frac{2}{\sqrt{-g}}\frac{\delta(\sqrt{-g})\mathcal{L}_m} {\delta g^{\mu\nu}}$ is the stress-energy tensor, $R_{\mu\nu}$ is the Ricci tensor, and $\square=g^{\mu\nu}\nabla_{\mu}\nabla_{\nu}$ is the d'Alembert operator. 

In $f(\mathcal{R})$ gravity, the field equations can be rearranged in terms of the Einstein tensor, comprising the higher-order curvature contributions
\begin{equation}
\label{ET}
    G_{\mu\nu}=\frac{1}{f_{\mathcal{R}}}\left(T_{\mu\nu}+T^{eff}_{\mu\nu}\right),
\end{equation}
where 
\begin{equation}
    T^{eff}_{\mu\nu}=\frac{f-\mathcal{R}f_{\mathcal{R}}}{2}g_{\mu\nu}+\left(\nabla_{\mu}\nabla_{\nu}-g_{\mu\nu}\square\right)f_{\mathcal{R}},
\end{equation}
where the effective energy-momentum tensor $T^{eff}_{\mu\nu}$ behaves as a source for the resulting geometry. This source term is usually referred to as the curvature fluid. Henceforth, when the stress-energy tensor for the curvature fluid is exactly zero, we recover GR. The difference between this equation and Einstein's equations, at least technically, is that the presence of $f$ denotes a non-minimal coupling. In this formulation, the effective gravitational coupling will not be a constant. 
To ensure the classical and quantum stability of a theory in the physically relevant domain, the following requirements on the derivatives of $f$ are crucial, that is, $f_{\mathcal{R}}>0$ and $f_{\mathcal{RR}} >0$. Indeed, the two conditions guarantee that gravity is attractive and that the effective gravitational constant is positive, respectively. It also ensures that the graviton is not a ghost and ensures the avoidance of the Dolgov-Kawasaki instability \cite{Appleby,Dolgov}.


\section{Raychaudhuri equations}
\label{section III}
In this section, we study the positive contributions of the Raychaudhuri equations for timelike geodesics, which guarantee the attractive character of the gravitational interaction in $f(\mathcal{R})$ theories. For a timelike congruence with velocity vector $u^{\mu}$, the Raychaudhuri equation is written as \cite{Raychaudhuri}
\begin{equation}
\label{R1}
    \frac{d\theta}{d\tau}=-\frac{1}{3}\theta^2+\nabla_{\mu}a^{\mu}-\sigma_{\mu\nu}\sigma^{\mu\nu}+w_{\mu\nu}w^{\mu\nu}-R_{\mu\nu}u^{\mu}u^{\nu},
\end{equation}
where $\tau$ is the affine parameter, $\theta=\nabla_{\mu}u^{\mu}$ is the expansion scalar, $u_{\mu}$ is the timelike velocity vector, $\sigma_{\mu\nu}=\nabla_{(_\nu u_\mu)}-\frac{1}{3}h_{\mu\nu}\,\theta+a_{(_\nu u_\mu)}$ is the shear, $h_{\mu\nu}$ is the spatial metric, $w_{\mu\nu}=\nabla_{[_\nu u_\mu]}-a_{[_\nu u_\mu]}$ is the rotation of a congruence of timelike geodesics, and $a^{\mu}=u^{\nu}\nabla_{\nu}\,u^{\mu}$ is the acceleration vector.

The last term on the right side of the Raychaudhuri equation, i.e., $-R_{\mu\nu}u^{\mu}u^{\nu}$, emphasizes the role of space-time geometry and is independent of the vector field's derivative. Using the field equations \eqref{fe} for $f(\mathcal{R})$ theory, this term may be expressed as follows:
\begin{equation}
 R_{\mu\nu}u^{\mu}u^{\nu}=\frac{1}{f_{\mathcal{R}}}\left(T_{\mu\nu}+\frac{f}{2}g_{\mu\nu}+\left(\nabla_{\mu}\nabla_{\nu}-g_{\mu\nu}\square\right)f_{\mathcal{R}}\right)u^{\mu}u^{\nu}.
\end{equation}
Consequently, this term has more general importance than the further terms in \eqref{R1}. This term may be understood geometrically as a mean curvature in the direction of $u$. The Raychaudhuri equation becomes more straightforward if we make the following assumptions: (i) Geodesics are the congruence of time-like curves. Then, $a_{\mu}=0$ is found along the geodesics. (ii) Timelike geodesic congruence to be hyper-surface orthogonal entails zero vorticity, i.e., the fourth term of the right-hand side of \eqref{R1} vanishes by virtue of the Frobenius Theorem. The Raychaudhuri equation \eqref{R1} becomes
\begin{equation}
     \frac{d\theta}{d\tau}=-\frac{1}{3}\theta^2-\sigma_{\mu\nu}\sigma^{\mu\nu}-R_{\mu\nu}u^{\mu}u^{\nu}.
\end{equation}

We now assume the Friedmann-Lemaitre-Robertson-Walker (FLRW) spacetime with a metric of the type that is homogeneous, isotropic, and spatially flat at the background level
\begin{equation}
\label{flrw}
ds^2=-dt^2+a^2(t)(dx^2+dy^2+dz^2),
\end{equation}
where $a(t)$ is the scale factor that is a function of cosmic time $t$. 
For such a metric and a matter distribution of a perfect fluid, given 
by $T_{\mu\nu}=(p+\rho)u_{\mu}u_{\nu}+p\,g_{\mu\nu}$, the Raychaudhuri equation \eqref{R1} takes the form,
\begin{equation}
    \label{RE}
    \frac{\Ddot{a}}{a}=\frac{1}{f_{\mathcal{R}}}\left(\frac{f(\mathcal{R})}{6}+H\,\Dot{\mathcal{R}}f_{\mathcal{R}\mathcal{R}}-\frac{\rho}{3}\right).
\end{equation}
So far, we have refrained from making any presumptions regarding the equation of state for the fluid distribution. Instead, the Raychaudhuri equation eliminates the fluid pressure $p$ using field equations \eqref{ET}.


\section{Bouncing solution in $f(\mathcal{R})$ gravity}
\label{section IV}
In this section, we examine the feasibility of getting suitable gravitational Lagrangians $f(\mathcal{R})$ capable of reproducing the cosmic evolution given by super-bounce and oscillatory bounce models. For a review of these bounces, one can check References \cite{Maria/2020}.

\subsection{Super-bounce}
The super-bouncing cosmology is distinguished by a power-law scale factor, which is first explored in \cite{Koehn/2014}. This bouncing model is used to construct a universe that collapses and rebirths without a singularity \cite{Oikonomou/2015}.
The Super-bounce scale factor and corresponding Hubble parameter are written as 
\begin{equation}
\label{1}
    a(t)\sim (t_s-t)^{2/c^2},\,\,\,\,\,\,\,H(t)=-\frac{2}{c^2(t_s-t)},
\end{equation}
where $t_s$ represents the time at which the bounce occurs.\\
Thus, we express the quantity $\frac{\Ddot{a}}{a}$ and the Ricci scalar $\mathcal{R}$ as
\begin{eqnarray}
    \frac{\Ddot{a}(t)}{a(t)}=\frac{2}{c^2}\left(\frac{2}{c^2}-1\right)\frac{1}{(t_s-t)^2},\\
\label{3}
    \mathcal{R}=6\frac{\Dot{a}^2}{a^2}+6\frac{\Ddot{a}}{a}=\frac{12}{c^2}\left(\frac{4}{c^2}-1\right)\frac{1}{(t_s-t)^2}.
\end{eqnarray}
Further, rewriting the equation \eqref{3} gives
\begin{equation}
    \frac{1}{(t_s-t)^2}=\frac{c^4}{12(4-c^2)}\mathcal{R}.
\end{equation}
By using the above expressions, the equation \eqref{RE} gives the following expression
\begin{equation}
\label{6}
    \mathcal{R}^2 f_{\mathcal{R}\mathcal{R}}+\frac{2-c^2}{2c^2}\mathcal{R}f_{\mathcal{R}}-\frac{4-c^2}{2c^2}f=-\frac{4-c^2}{c^2}\rho.
\end{equation}
Assuming the barotropic equation of state $p=\omega\rho$ yields the energy density using the conservation equation as $\rho=\rho_0\,a^{-3(1+\omega)}$. Henceforth, we obtain the energy density in terms of $\mathcal{R}$ (using Eqs. \eqref{1} and \eqref{3}) given by
\begin{equation}
    \label{rho1}
    \rho=\rho_0\,\left(\frac{c^4\,\mathcal{R}}{12(4-c^2)}\right)^{\frac{3(1+\omega)}{c^2}}.
\end{equation}
Consequently, the equation \eqref{6} takes the form 
\begin{equation}
    \label{DE1}
    \mathcal{R}^2 f_{\mathcal{R}\mathcal{R}}+\frac{2-c^2}{2c^2}\mathcal{R}f_{\mathcal{R}}-\frac{4-c^2}{2c^2}f=\mathcal{X}_1\,\mathcal{R}^{\gamma},
\end{equation}
where $\mathcal{X}_1=-\frac{\rho_0\,(4-c^2)}{c^2}\left(\frac{c^4}{12(4-c^2)}\right)^{\frac{3(1+\omega)}{c^2}}$.\\
In fact, one can now obtain the solution to the above differential equation given by
\begin{equation}
    f(\mathcal{R})=c_1\,\mathcal{R}^{\alpha}+c_2\,\mathcal{R}^{\beta}+\mathcal{X}_2\,\mathcal{R}^{\gamma},
\end{equation}
where $c_1$ and $c_2$ are integrating constants and
\begin{eqnarray}
\nonumber
   && \alpha=-\frac{2-3c^2+\sqrt{4+20c^2+c^4}}{4c^2},\\
    \nonumber
    && \beta=\frac{-2+3c^2+\sqrt{4+20c^2+c^4}}{4c^2},\\
    \nonumber
   && \gamma=\frac{3(1+\omega)}{c^2},\\
    \nonumber
   && \mathcal{X}_2=\frac{2\mathcal{X}_1\,c^4}{c^4-c^2(13+9\omega)+6(4+7\omega+3\omega^2)}.
\end{eqnarray}

\begin{figure}[H]
\includegraphics[scale=0.6]{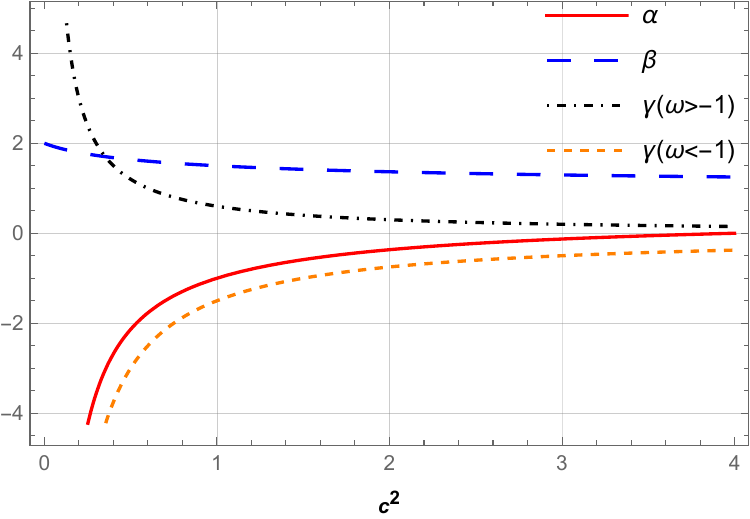}
\caption{Variation of model parameters $\alpha$, $\beta$, $\gamma$ with respect to $c^2$.}
\label{fig1}
\end{figure}
One can notice that the solution for $f(\mathcal{R})$ incorporates three distinct powers of $\mathcal{R}$. The behavior of $\alpha$, $\beta$, and $\gamma$ with respect to $c^2$ is depicted in FIG. \ref{fig1}. Notably, for $0<c^2<4$, $\alpha$ consistently is negative, while $\beta$ remains positive. However, the value of $\gamma$ is contingent upon the equation of state parameter $\omega$. In regions where $\omega > -1$ (non-phantom phase), $\gamma$ invariably assumes positive values, whereas in the phantom phase, where $\omega < -1$, $\gamma$ consistently adopts negative values.

\subsubsection{Strong Energy Condition}
The possibility of accelerating expansion without the requirement for dark energy (or an inflaton field) is a significant part of the motivation for $f(\mathcal{R})$ gravity. A straightforward way to see this is to define the effective pressure and effective energy density of the geometry as \cite{Sotiriou}
\begin{eqnarray}
   && H^2=\frac{\rho_{eff}}{3},\\
   && \frac{\Ddot{a}}{a}=-\frac{\rho_{eff}+3p_{eff}}{6},
\end{eqnarray}
where $p_{eff}$ and $\rho_{eff}$ are the effective
pressure and energy density, respectively.

In this context, we may also examine the effective energy condition by calculating the quantity $\rho_{eff}+3p_{eff}$, which in this case is given by
\begin{equation}
    \rho_{eff}+3p_{eff}=\frac{12}{c^4}(c^2-4)\frac{1}{(t_s-t)^2}.
\end{equation}
The negative nature of the solution for $f(\mathcal{R})$ within the range $0<c^2<4$ signifies a violation of the strong energy condition. This aligns with expectations for an eternally accelerated model.


\subsubsection{Viability Criteria}
\begin{figure}[]
\includegraphics[scale=0.63]{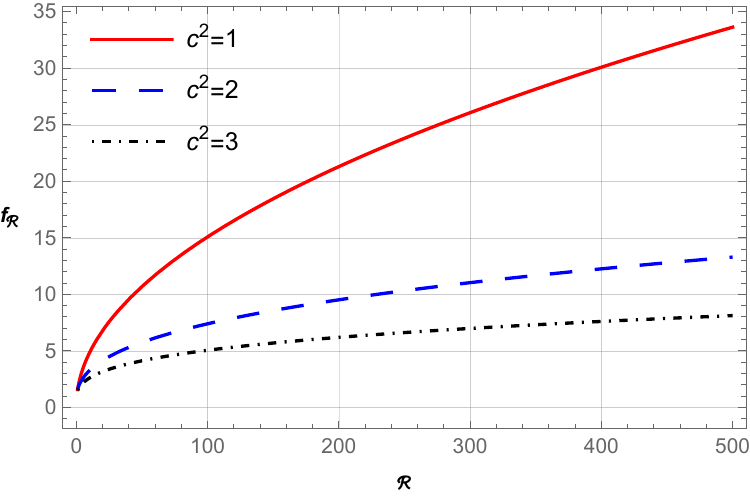}  
\includegraphics[scale=0.63]{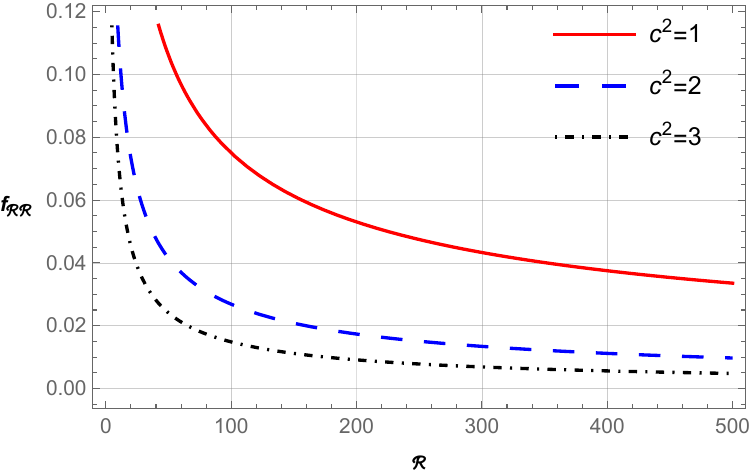}
\caption{Behavior of $f_{\mathcal{R}}$ and $f_{\mathcal{R}\mathcal{R}}$ vs. $\mathcal{R}$ for different values of $c^2$.}
\label{fig23}
\end{figure}

We aim to verify the theoretical consistency of the relevant $f(\mathcal{R})$ gravity model. Any viable $f(\mathcal{R})$ model must adhere to the conditions $f_{\mathcal{R}}>0$ and $f_{\mathcal{R}\mathcal{R}}>0$ \cite{Sawicki/2007,Silvestri/2009,Sotiriou}.
The stability of the model hinges on the latter requirement, while the former ensures the positivity of the effective gravitational constant. Our obtained model satisfies both conditions $f_{\mathcal{R}}>0$ and $f_{\mathcal{R}\mathcal{R}}>0$ (evidenced in FIG. \ref{fig23}). 

We find the viability of the model at low and high curvature for various phases as follows:
\begin{itemize}
    \item \textit{For non-phantom phase :-} In the expression of $f(\mathcal{R})$, the first term dominates and $\mathcal{R}$ approaches $0$ as $\alpha$ is negative. Thus, for the model to be valid at low curvature, it is necessary to have $c_1=0$, ensuring the viability. Additionally, the model remains viable at high curvatures.
    
    \item \textit{For phantom phase :-} In this case, the dominance shifts towards the first and third terms as $\mathcal{R}\to 0$ due to negative values of $\alpha$ and $\gamma$. As a result, for the model to be viable at low curvature, we must have $c_1=0$, and $\mathcal{X}_2=0$. Further, the model remains viable when curvature is high.
    \item \textit{For $\Lambda$CDM ($\omega=-1$) :-} According to the functional form obtained for $f(\mathcal{R})$, the third term remains constant as $\mathcal{X}_2=\frac{2\rho_0}{c^2}$. Further, $c_1=0$ satisfies the viability criteria at low curvature. This simplifies our model to $f(\mathcal{R})=c_2\,\mathcal{R}^{\beta}-2\Lambda(=\mathcal{X}_2)$. Additionally, in FIG. \ref{fig1}, as $c^2$ increases, the parameter $\beta$ tend towards $1$, aligning our model with the $\Lambda$CDM ($f(\mathcal{R})=\mathcal{R}-2\Lambda$), ensuring consistency with the current standard model of cosmology that fits most large-scale observational data.    
\end{itemize}

\subsection{Oscillatory bounce}
A study on an oscillatory bouncing model in cosmology has gained attention for its potential advantages and alternative approach to understanding the universe. The Oscillatory bouncing cosmology is distinguished by a periodic scale factor and the corresponding Hubble parameter, given by 
\begin{equation}
\label{10}
    a(t)=A\,\sin^2\left(\frac{Bt}{t_*}\right),\,\,\,\,\,\,\,H(t)=\frac{2B}{t_*}\,\cot\left(\frac{Bt}{t_*}\right).
\end{equation}
where $t_*>0$ is some reference time, $A>0$ and $B>0$ are dimensionless constants. This bouncing model characterizes a cyclic universe as a series of contractions and expansions (periodically).
With a choice of scale factor, the model displays two different bouncing patterns. Each cycle begins with a ``Big Bang", finishes with a "big crunch", and then begins again with a ``Big Bang". 

The first scenario involves a singularity occurring regularly within each cycle, characterized by the scale factor reaching zero and the Hubble parameter becoming singular, indicating a Big Crunch or Big Bang singularity. The second bounce occurs when the universe reaches its maximum considerable extent, ceasing expansion and initiating its contracting toward Big Crunch singularity. For more details, one can review these References \cite{Tolman,Steinhardt/2002,Khoury/2004,Maria/2020}.

We define the quantity $\frac{\Ddot{a}}{a}$ and the Ricci scalar $\mathcal{R}$ as
\begin{eqnarray}
  &&  \frac{\Ddot{a}}{a}=\frac{2B^2}{t^2_*}\left[\cot^2\left(\frac{Bt}{t_*}\right)-1\right],\\
\label{12}
   && \mathcal{R}=6\frac{\Dot{a}^2}{a^2}+6\frac{\Ddot{a}}{a}=\frac{12B^2}{t^2_*}\left[3\,\cot^2\left(\frac{Bt}{t_*}\right)-1\right].
\end{eqnarray}
One can rewrite Eq. \eqref{10} as
\begin{equation}
\label{13}
    \sin^2\left(\frac{Bt}{t_*}\right)=\frac{a}{A},\,\,\,\,\,\cot^2\left(\frac{Bt}{t_*}\right)=\frac{A}{a}-1.
\end{equation}
Furthermore, using Eqs.\eqref{12} and \eqref{13}, we obtain 
\begin{equation}
\label{14}
    \frac{A}{a}=\frac{t_*^2}{36\,B^2}\mathcal{R}+\frac{4}{3}.
\end{equation}
Hence, introducing all the above expressions, Eq. \eqref{RE} turns out to be 
\begin{multline}
\label{15}
      \left(\mathcal{R}+\frac{12\,B^2}{t_*^2}\right)\left(\mathcal{R}+\frac{48\,B^2}{t_*^2}\right)\,f_{\mathcal{R}\mathcal{R}}\\+\frac{1}{2}\left(\mathcal{R}-\frac{24\,B^2}{t_*^2}\right)\,f_{\mathcal{R}}-\frac{3}{2}\,f=-3\,\rho.
\end{multline}
We further use the barotropic equation of state $p=\omega\rho$ to obtain the energy density as $\rho=\rho_0\,a^{-3(1+\omega)}$. Now, by using Eqs.\eqref{1} and \eqref{3}, we express the energy density in terms of $\mathcal{R}$ as
\begin{equation}
    \label{rho1}
    \rho=\rho_0\left(\frac{t_*^2}{36A\,B^2}\right)^{3(1+\omega)}\,\left(\mathcal{R}+\frac{48\,B^2}{t_*^2}\right)^{3(1+\omega)}.
\end{equation}
The equation \eqref{15} takes the form 
\begin{multline}
\label{RR}
      \left(\mathcal{R}+\frac{12\,B^2}{t_*^2}\right)\left(\mathcal{R}+\frac{48\,B^2}{t_*^2}\right)\,f_{\mathcal{R}\mathcal{R}}+\\ \frac{1}{2}\left(\mathcal{R}-\frac{24\,B^2}{t_*^2}\right)\,f_{\mathcal{R}}-\frac{3}{2}\,f=\mathcal{K}\,\left(\mathcal{R}+\frac{48\,B^2}{t_*^2}\right)^{3(1+\omega)},
\end{multline}
where $\mathcal{K}=-3\rho_0\left(\frac{t_*^2}{36A\,B^2}\right)^{3(1+\omega)}$.\\
To simplify the above differential equation, we put $\lambda=\frac{12B^2}{t_*^2}$ to obtain 
\begin{multline}
\label{18}
    (\mathcal{R}+\lambda)(\mathcal{R}+4\lambda)\,f_{\mathcal{R}\mathcal{R}}+\frac{1}{2}(\mathcal{R}-2\lambda)\,f_{\mathcal{R}}-\frac{3}{2}\,f\\=\mathcal{K}\,\left(\mathcal{R}+4\lambda\right)^{3(1+\omega)}.
\end{multline}
The corresponding homogeneous equation above can be transformed into the hypergeometric differential equation by the substitution $x=-\frac{1}{3\lambda}(\mathcal{R}+\lambda)$, as
\begin{equation}
    x(1-x)\frac{d^2f}{dx^2}-\frac{1}{2}(x+1)\frac{df}{dx}+\frac{3}{2}f=0.
\end{equation}
The solution of the homogeneous differential equation is
\begin{equation}
    f_H(x)=c_1\,_2F_1\left(1,-\frac{3}{2},-\frac{1}{2},x\right)+c_2\,x^{3/2},
\end{equation}
where $_2F_1$ is a hypergeometric function. To find the particular solutions, we categorize them based on the EoS parameter $\omega$. In these cases, we observe that the real-valued particular solutions are only obtained when $\omega>-1$ (non-phantom phase), while the solutions become imaginary when $\omega<-1$ (phantom phase).

\begin{widetext}

\begin{table}[]
\begin{center}
  \caption{The particular integral for differential Eq. \eqref{18} for different values of the EoS parameter $\omega$.}
    \label{Table}
    \begin{tabular}{l c c c c}
\hline\hline 
3$(1+\omega)$          & EoS parameter & Particular Integral &  solutions     \\ \hline\hline 
    $3(1+\omega)=0$       & $\omega=-1$ &   $f_p(\mathcal{R})=2\rho_0(=2\Lambda)$ & Accepted\\[1ex]
$3(1+\omega)=1$           & $\omega=-\frac{2}{3}$ & $f_p(\mathcal{R})=-\mathcal{K}(\mathcal{R}+2\lambda)$ & Accepted \\[1ex] 
$3(1+\omega)=2$           & $\omega=-\frac{1}{3}$ & $f_p(\mathcal{R})=\mathcal{K}\left(6\mathcal{R}^2-\frac{200}{27}\lambda\,\mathcal{R}-16\lambda^2\right)-\frac{128\,\mathcal{K}}{81}\sqrt{3\lambda}\,(\mathcal{R}+\lambda)^{3/2}\, \tan^{-1}\left(\sqrt{\frac{\mathcal{R}+\lambda}{3\lambda}}\right)$ & Accepted \\[1ex]
$3(1+\omega)<0$           & $\omega<-1$ & Imaginary solutions & Rejected \\[1ex] \hline
\end{tabular}
\end{center}
\end{table}
\end{widetext}

\subsubsection{Strong Energy Condition}
We can additionally analyze the effective energy condition to assess an oscillatory bounce. This entails evaluating the quantity $\rho_{eff}+3p_{eff}$, which for this case is 
\begin{equation}
    \rho_{eff}+3p_{eff}=-\frac{12B^2}{t_*}\left[\cot^2\left(\frac{B\,t}{t_*}\right)-1\right].
\end{equation}
The expression above is negative for $\cot^2\left(\frac{B\,t}{t_*}\right)>1$, i.e., $-\frac{(2n+1)\pi}{4}<\frac{B\,t}{t_*}<\frac{(2n+1)\pi}{4}$ and thus violates the strong energy condition, which is expected for an eternally accelerated model. 

\subsubsection{ Viability Criteria}
\begin{widetext}

\begin{figure}[H]
\centering
\subcaptionbox{}
{\includegraphics[width=0.31\textwidth]{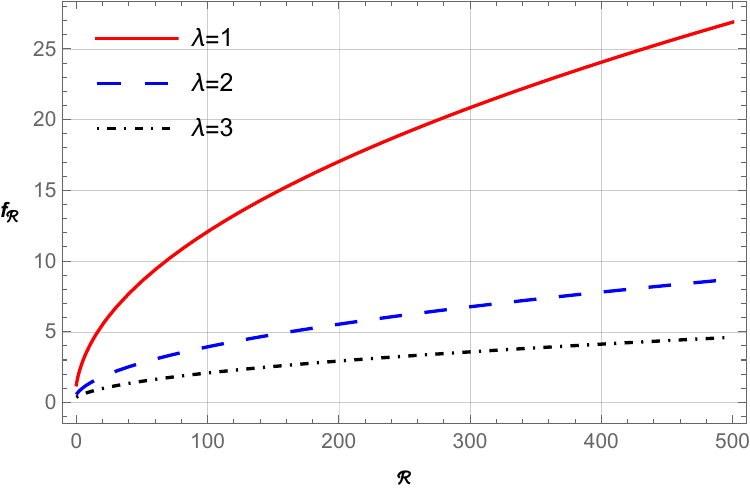}}
 \hspace{0.15in} 
\subcaptionbox{}{\includegraphics[width=0.31\textwidth]{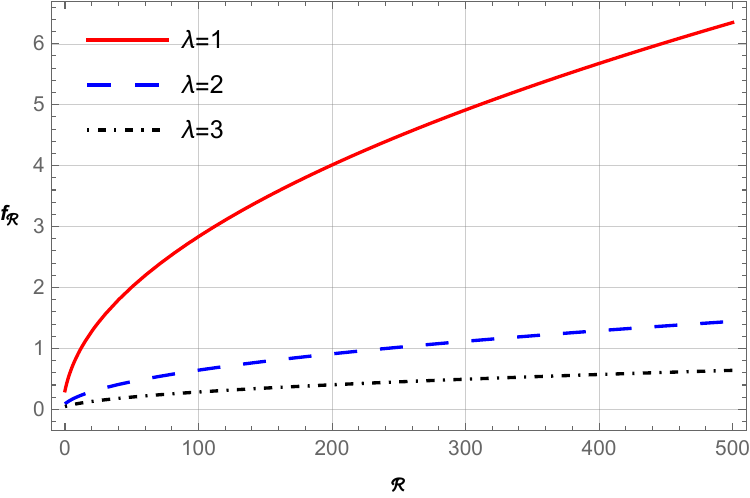}}%
 \hspace{0.15in} 
\subcaptionbox{}{\includegraphics[width=0.31\textwidth]{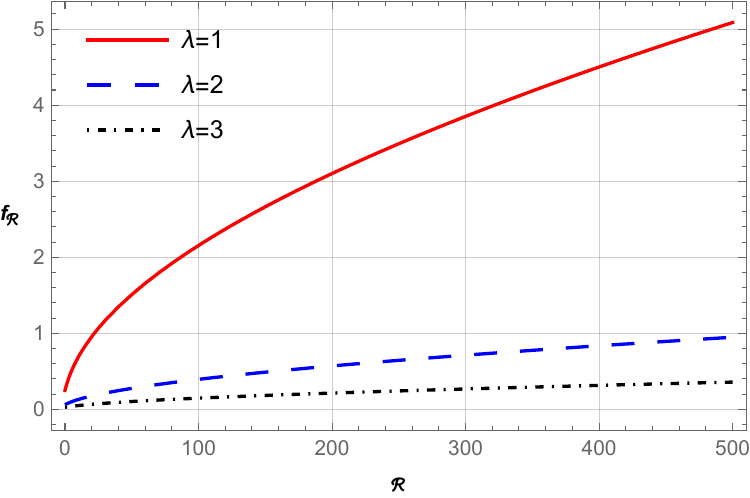}}%
\hspace{0.15in} 
\subcaptionbox{}{\includegraphics[width=0.31\textwidth]{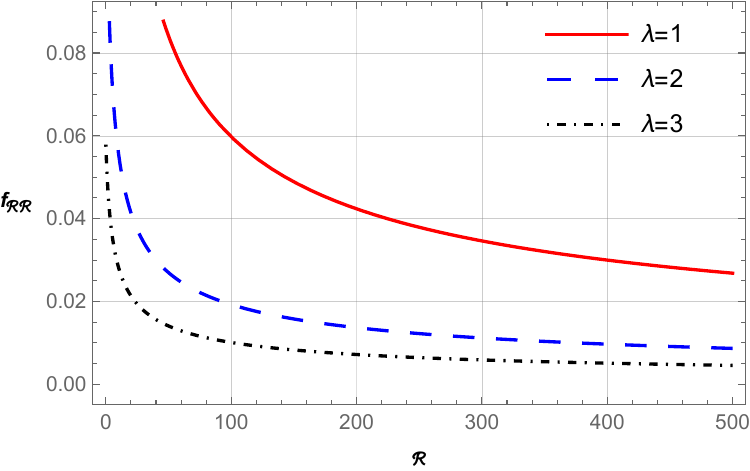}}%
\hspace{0.15in} 
\subcaptionbox{}{\includegraphics[width=0.31\textwidth]{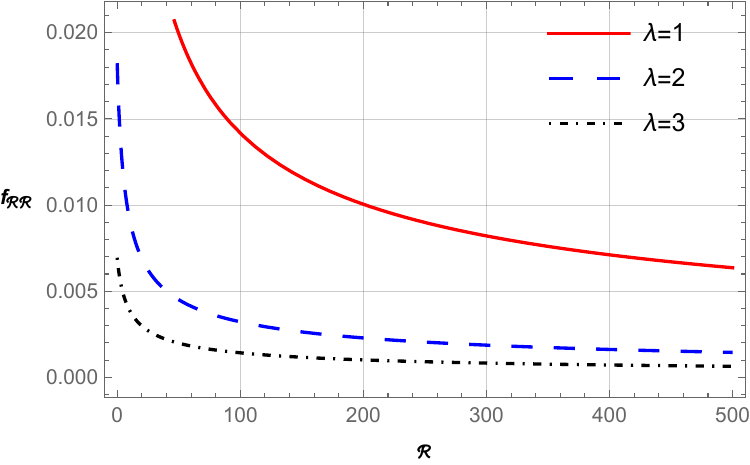}}%
\hspace{0.15in} 
\subcaptionbox{}{\includegraphics[width=0.31\textwidth]{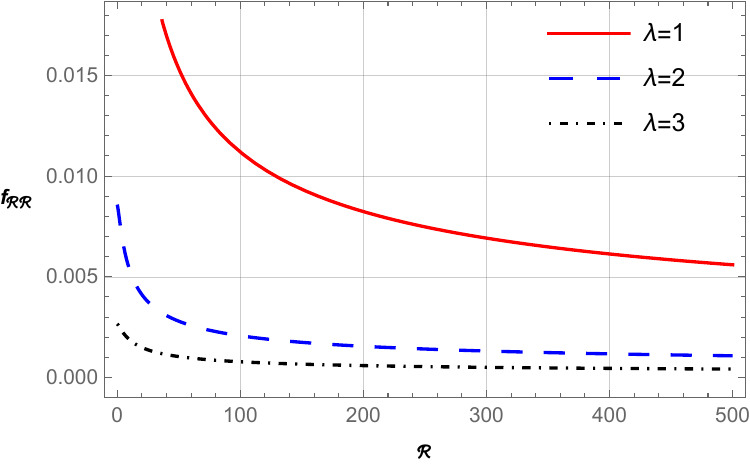}}%
\\ 
\caption{Plots (a) and (d) illustrate the behavior of $f_{\mathcal{R}}$ and $f_{\mathcal{R}\mathcal{R}}$ for $\omega=-1$. Plots (b) and (e) depict the behavior of $f_{\mathcal{R}}$ and $f_{\mathcal{R}\mathcal{R}}$ for $\omega=-2/3$. Plots (c) and (f) exhibit the behavior of $f_{\mathcal{R}}$ and $f_{\mathcal{R}\mathcal{R}}$  for $\omega=-1/3$. }
\label{fig4}
\end{figure}

\end{widetext}

We seek to verify the theoretical consistency of the relevant $f(\mathcal{R})$ gravity model. Our derived models in all instances satisfy both the conditions $f_{\mathcal{R}}>0$ and $f_{\mathcal{R}\mathcal{R}}>0$, as depicted in Figs. \ref{fig4}). We assess the viability of the model at low and high curvature regimes for various phases as follows:
\begin{itemize}
    \item \textit{For $3(1+\omega)=0$ :-} For this scenario, $f(R)\to 0$ as $R\to 0$ when we have $c_1 \lambda ^{3/2}+\frac{\pi\,c_2}{27 \sqrt{3} \lambda }=\frac{2\rho_0}{3}$. As a result, at low curvature, we must have $c_1=\frac{54 \lambda\,\rho_0-\sqrt{3} \pi\, c_2}{81 \lambda ^{5/2}}$ to meet the viability criterion, and hence, the model will consistently remain viable at large curvature.

    \item \textit{For $3(1+\omega)=1$ :-} Here, we obtain that, $f(R)\to 0$ as $R\to 0$ when $c_1\,\lambda^{3/2}+\frac{\pi\,c_2}{27\sqrt{3} \lambda}=2\lambda\mathcal{K}$. Consequently, we must have $c_1=\frac{162 \mathcal{K} \lambda ^2-\sqrt{3} \pi\, c_2}{81 \lambda ^{5/2}}$ at low curvature to be viable. Similarly, the model is viable at large curvature also.
    
   \item \textit{For $3(1+\omega)=2$ :-} In this scenario, $f(R)\to 0$ as $R\to 0$ for $c_1 \lambda ^{3/2}-\frac{\pi  \left(64 \lambda^3\,\mathcal{K} -3\,c_2\right)}{81\,\sqrt{3} \lambda }=\frac{16}{3}\lambda^2\,\mathcal{K}$. So, at low curvature, we get $c_1=\frac{-3 \pi  \sqrt{3}\,c_2+1296 \mathcal{K} \lambda^3+ 64 \pi  \sqrt{3} \mathcal{K} \lambda ^3}{243 \lambda ^{5/2}}$ for the viability requirement. However, the model will consistently remain viable at high curvature.
\end{itemize}








\section{Cosmological aspects of reconstructed models}
\label{section v}
Returning to our discussion, assuming that the stress-energy tensor is $T_{\mu\nu}=(p+\rho)u_{\mu}u_{\nu}+p\,g_{\mu\nu}$, and inserting the flat FLRW metric \eqref{flrw} in the field equations \eqref{fe} yields \cite{Sotiriou,Santos}
\begin{eqnarray}
3H^2&=&\frac{\kappa}{f_{\mathcal{R}}}\left[\rho+\frac{Rf_{\mathcal{R}}-f}{2}-3H\Dot{\mathcal{R}}f_{\mathcal{R}\mathcal{R}}\right],\\
   2\Dot{H}+3H^2&=&-\frac{\kappa}{f_{\mathcal{R}}}\left[p+\Dot{\mathcal{R}}^2f_{\mathcal{R}\mathcal{R}\mathcal{R}}+2H\Dot{\mathcal{R}}f_{\mathcal{R}\mathcal{R}}\right.\\
 && \left.+\Ddot{\mathcal{R}}f_{\mathcal{R}\mathcal{R}}+\frac{1}{2}(f-\mathcal{R}f_{\mathcal{R}})\right].
\end{eqnarray}
An important part of the motivation for $f(\mathcal{R})$ gravity is that it can result in accelerated expansion without the need for dark energy. To exemplify this, let us define the effective energy density and pressure as follows
\begin{eqnarray} 
\nonumber
    \rho_{eff}&=&\frac{Rf_{\mathcal{R}}-f}{2f_{\mathcal{R}}}-\frac{3H\Dot{\mathcal{R}}f_{\mathcal{R}\mathcal{R}}}{f_{\mathcal{R}}}, \\ \nonumber
p_{eff}&=&\frac{\Dot{\mathcal{R}}^2f_{\mathcal{R}\mathcal{R}\mathcal{R}}+2H\Dot{\mathcal{R}}f_{\mathcal{R}\mathcal{R}}+\Ddot{\mathcal{R}}f_{\mathcal{R}\mathcal{R}}+\frac{1}{2}(f-\mathcal{R}f_{\mathcal{R}})}{f_{\mathcal{R}}},
\end{eqnarray}
where $\rho_{eff}$ has to be non-negative in a spatially flat FLRW spacetime.

The effective equation of state parameter $\omega_{eff}=\frac{p_{eff}}{\rho_{eff}}$ for $f(\mathcal{R})$ can be represented as
\begin{equation}
\omega_{eff}=\frac{\Dot{\mathcal{R}}^2f_{\mathcal{R}\mathcal{R}\mathcal{R}}+2H\Dot{\mathcal{R}}f_{\mathcal{R}\mathcal{R}}+\Ddot{\mathcal{R}}f_{\mathcal{R}\mathcal{R}}+\frac{1}{2}(f-\mathcal{R}f_{\mathcal{R}})}{\frac{Rf_{\mathcal{R}}-f}{2}-3H\Dot{\mathcal{R}}f_{\mathcal{R}\mathcal{R}}}.
\end{equation}

Energy conditions (ECs) serve as fundamental tools for understanding the geodesics of the universe. These conditions can be derived using the well-known Raychaudhury equations, as described in Eq. \eqref{R1}. The ECs are characterized as \cite{Santos,Carroll/2004}
\begin{itemize}
	\item Strong Energy Conditions (SEC) :  $\rho_{eff} + 3p_{eff} \geq 0$
	\item Weak Energy Conditions (WEC) : $\rho_{eff}\geq 0$,\\   $p_{eff} + \rho_{eff} \geq 0$
	\item Null Energy Conditions (NEC) : $p_{eff} + \rho_{eff} \geq 0$
	\item Dominant Energy Conditions (DEC) : If $\rho_{eff}\geq 0$,   $\rvert p_{eff} \rvert \leq \rho_{eff} $.
\end{itemize}
One idea is to let us think about the singularity-free universal beginning, which is the universal bouncing scenario. Many suggestions in the literature suggested avoiding this singularity via quantum features, but these lack the dependability to integrate into gravitational theory. So, at this stage, gravitational theories provide a specific technique for testing the validity of both the bounce model and its own. The null energy condition (NEC) is a technique for accomplishing the objective. Furthermore, it has been demonstrated that in the setting of GR, violating NEC is exceedingly difficult for local-field models. As a result, effective field theories allow for the detection of NEC violations and the possibility of a non-singular bounce \cite{NEC}. However, in standard general relativity, the strong energy condition is closely tied to the idea that gravity is always attractive. If this condition is violated in modified gravity, it could lead to repulsive gravitational effects, potentially contributing to accelerated cosmic expansion. The dominant energy conditions are linked to the stability of solutions in gravitational theories, and its validation suggests that the spacetime configurations and solutions remain stable and do not exhibit instabilities that could lead to unexpected behavior or violations of causality.

\begin{figure}[H]
\includegraphics[scale=0.63]{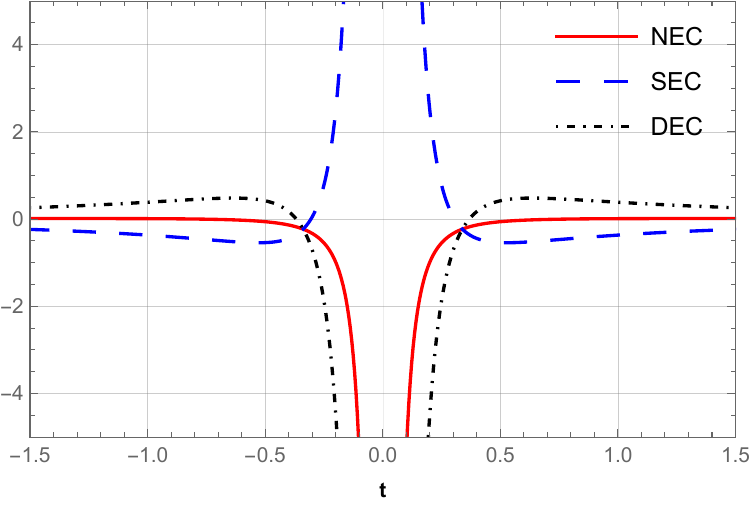}
\caption{Energy conditions with respect to time using $\beta=0.98$, $\gamma=0.85$ ($\gamma>0$ represent the non-phantom case), $c_2=-1.75$ and $\mathcal{X}_2=0.5$ for the Super-bounce model with $0<c^2<4$.}
\label{EC1}
\end{figure}

\begin{figure}[H]
\includegraphics[scale=0.63]{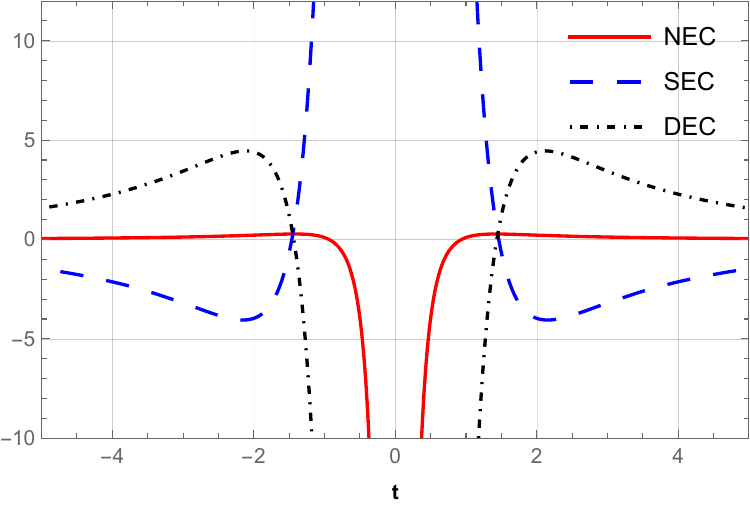}
\caption{Energy conditions with respect to time using $\beta=0.2$, $\gamma=-0.7$ ($\gamma<0$ represent the phantom case), and $c_2=-0.75$ for the Super-bounce model with $0<c^2<4$.}
\label{EC2}
\end{figure}
We test our reconstructed models through the energy conditions. The energy conditions for the reconstructed from super-bounce are displayed in Figs. \ref{EC1} (non-phantom case) and \ref{EC2} (phantom case). In both figures, we can see that the NEC is violated near the bouncing point and satisfied after a certain point. The violation of NEC near the singular point derives the bouncing nature of the universe. Also, the SEC is violated after a certain point toward the late-time universe, and this could mimic the effects of dark energy, the mysterious force believed to be driving the accelerated expansion of the universe. Furthermore, the DEC is satisfied after a certain point toward the late-time universe. Hence, we conclude that our reconstructed model is useful for studying the singular free accelerating universe. 

\begin{figure}[H]
\includegraphics[scale=0.65]{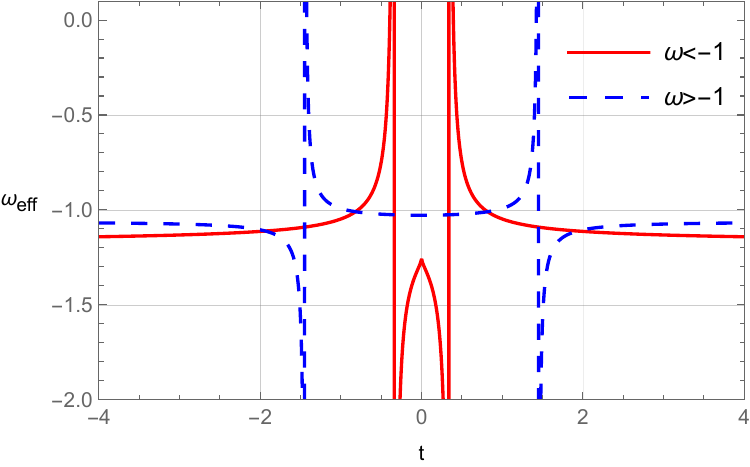}
\caption{Behavior of the effective EoS parameter versus cosmic time $t$ for the Super-bounce model.}
\label{w1}
\end{figure}

The effective equation of state parameter shows the phantom dark energy behavior after a certain point toward the late-time universe (see Fig. \ref{w1}) for the same value of the model parameters. Phantom dark energy, characterized by $\omega_{eff}<-1$, leads to accelerated cosmic expansion. The negative pressure associated with phantom dark energy causes a repulsive gravitational effect, driving the universe to expand at an increasing rate.

\begin{figure}[]
\includegraphics[scale=0.78]{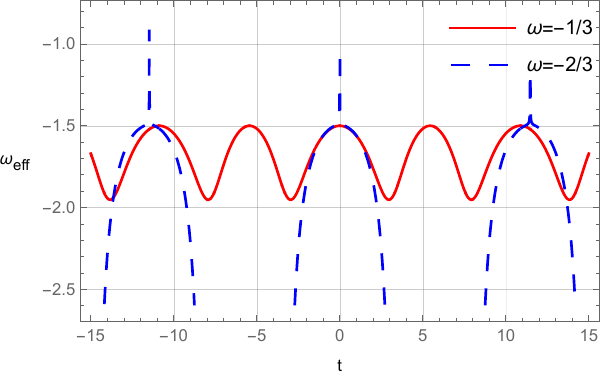}
\caption{Behavior of the effective EoS parameter with respect to time for the oscillatory bounce.}
\label{w2}
\end{figure}

\begin{figure}[]
\includegraphics[scale=0.75]{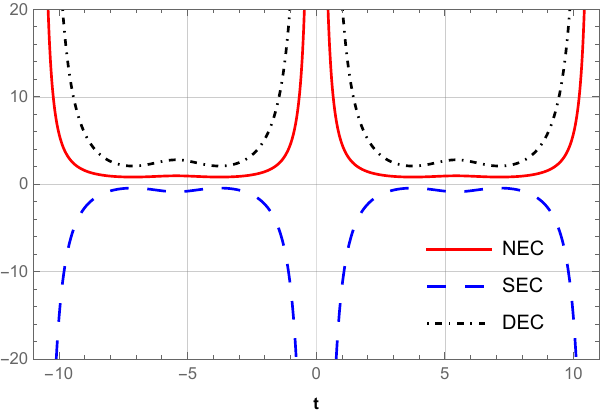}
\caption{Energy conditions with respect to time for the oscillatory bounce for $\omega=-2/3$ using the value of parameters $\lambda=1$ (or $2$ or $3$), $\mathcal{K}=1$, and $c_2=-1$. }
\label{EC3}
\end{figure}

\begin{figure}[]
\includegraphics[scale=0.75]{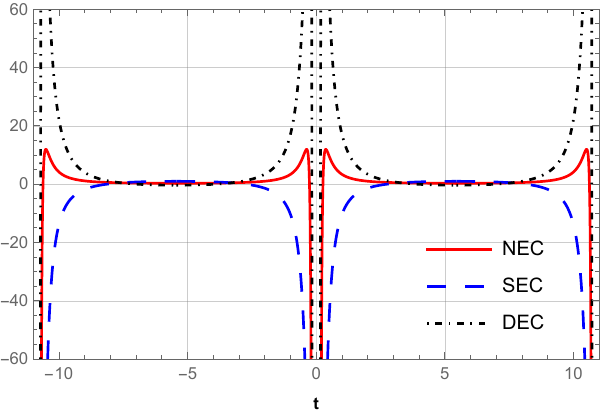}
\caption{Energy conditions with respect to time for the oscillatory bounce for $\omega=-1/3$ using the value of parameters $\lambda=1$ (or $2$ or $3$), $\mathcal{K}=-0.005$, and $c_2=-1$.}
\label{EC4}
\end{figure}

The energy conditions for the oscillatory bounce model are plotted in Figs. \ref{EC3} (for $\omega=-1/3$) and \ref{EC4} (for $\omega=-2/3$). In Fig. \ref{EC3}, only the strong energy condition (SEC) is violated, and the other two NEC and WEC are satisfied. Hence, the results of oscillatory bounce model for the case $\omega=-2/3$ tested through the energy conditions show the acceleration of the universe, but it fails to show the singular free universe. In Fig. \ref{EC4}, we can see that the NEC is violated near the bouncing point and satisfies its condition after a certain point. The violation of NEC near the singular point derives the bouncing nature of the universe. In this figure, the SEC is also violated, and it can show the negative pressure regime of the universe. Furthermore, the DEC is satisfied at late times. Hence, we conclude that our reconstructed oscillatory model for the case $\omega=-1/3$ is useful for studying the singular free accelerating universe. The effective equation of state parameter shows the phantom dark energy behavior (see Fig. \ref{w2}) for the same values of the model parameters.

\section{conclusion}
\label{section VI}

The conventional cosmological model posits that the universe originated from a singularity, commonly referred to as the Big Bang. Nonetheless, this model abounds with several challenges, including the horizon problem, flatness problem, original structure problem, entropy problem, and the persisting issue of a singularity. In order to surmount these obstacles, a rapid phase of expansion immediately following the Big Bang was deemed necessary to establish a universe characterized by uniformity, flatness, and smoothness. Addressing these concerns, Alan Guth, A.A. Starobinsky and Katsuhiko Sato \cite{Staro,Sato} formulated an inflationary theory, which has proven remarkably successful in explaining various observational features of the universe. Despite these challenges, bouncing cosmology offers a promising avenue for exploring alternative scenarios to the conventional Big Bang model and has the potential to address some of the longstanding puzzles in cosmology. Ongoing theoretical and observational efforts are aimed at further elucidating the feasibility and implications of bouncing cosmology in understanding the origin and evolution of the universe.

In this study, we have formulated $f(\mathcal{R})$ models specifically developed for addressing the challenge of the initial singularity in both super-bounce and oscillatory bounce scenarios. The construction of these $f(\mathcal{R})$ models is rooted in the Raychaudhari equation, a key element incorporated in reconstructed models designed to circumvent singularities. The solutions we derived are unique in the sense that we identified both the general and particular solutions for $f(\mathcal{R})$ models. Additionally, we found power-law and hypergeometric solutions for the cases considered, which align with solutions from other studies in the literature \cite{Gupta} on $f(\mathcal{R})$ theory. The reconstructed $f(\mathcal{R})$ is a combination of powers of $\mathcal{R}$, but potentially a wide variety of models can be found in the present work as the powers of $\mathcal{R}$ are not uniquely determined. In the case of the hypergeometric solutions, the model is valid for large curvature regimes to the present epoch (low curvature) subject to a tuning of the constants.  It is important to note that if metric anisotropy is considered an independent dynamic degree of freedom, the situation will differ. Interested readers can refer to \cite{Bhattacharya} for details on the novelties of such anisotropic dynamics in metric $f(\mathcal{R})$ gravity, where it was shown that, even for the simple case of $\mathcal{R}+\alpha\,\mathcal{R}^2$ gravity, there are generally three possible solutions for total anisotropy given an average scale factor. This can be explored in more detail in future studies.

Furthermore, we examined the feasibility of getting suitable gravitational Lagrangians for $f(\mathcal{R})$ capable of reproducing the cosmic evolution given by super-bounce and oscillatory bounce models.

\textit{Super-bounce :-} We have obtained the power-law type $f(\mathcal{R})$ model for this bouncing scenario. We have investigated this model for two phases: non-phantom and phantom. Our analysis of the super-bounce model involves evaluating the quantity $\rho_{eff}+3p_{eff}$, leading to the constraint $0<c^2<4$ (as expected for an accelerated model). We find the viability of the model at low and high curvature for both non-phantom and phantom cases, imposing certain constraints on model parameters. Furthermore, we have verified the viability of our obtained model through two conditions: firstly, ensuring $f_{\mathcal{R}}>0$ to maintain a positive effective gravitational constant, which also ensures that the graviton is not a ghost \cite{Appleby}, and secondly, confirming $f_{\mathcal{R}\mathcal{R}}>0$ to ensure stability, i.e., the avoidance of the Dolgov-Kawasaki instability.

\textit{Oscillatory bounce :-} In this analysis, we can see in Table \ref{Table} that our particular solution varies for different phases of the universe, suggesting the importance of employing corresponding models to study each phase effectively. We further investigate the effective strong energy condition by evaluating the quantity $\rho_{eff}+3p_{eff}$. The effective strong energy condition is violated when $\cot^2\left(\frac{B\,t}{t_*}\right)>1$, i.e., $-\frac{(2n+1)\pi}{4}<\frac{B\,t}{t_*}<\frac{(2n+1)\pi}{4}$, aligning with expectations for an endlessly accelerating model. Similar to the previous model, we found the viability of our model at low and high curvature and confirmed that the obtained model can be stable. 

Our reconstructed models underwent rigorous testing based on energy conditions, yielding noteworthy results. These include the avoidance of classical cosmological singularities linked to the Big Bang and the Big Crunch. Additionally, our model demonstrates the potential to drive the accelerated expansion of the universe. We observed the violation of NEC near the bouncing point, followed by the validation after a certain threshold. This violation of the NEC near the singularity underscored the bouncing nature of the universe. Moreover, the SEC is violated towards the late-time universe, potentially depicting the accelerated expansion. The validation of DEC indicated that the spacetime configurations and solutions maintain stability, avoiding instabilities that might result in unexpected behavior or violations of causality.  Furthermore, the effective EoS shows the negative pressure associated with phantom dark energy causes a repulsive gravitational effect, driving the universe to expand at an increasing rate. 
One can note that the $f(\mathcal{R})$ models seem viable and cannot be ruled out only assuming the mathematical equivalence. However, considering simple $f(R) =  f_{0} R^n$ models cannot completely solve the cosmic evolution problem since matter-dominated solutions for $n \neq 1$ are generally stable and do not give rise to late-time transitions to accelerated behaviors. The situation is similar to that faced several times in inflationary cosmology, where a graceful exit from the inflation regime can result in problems for simple models. In other words, $f(\mathcal{R})$ gravity could be the way to bypass shortcomings such as dark energy and dark matter at cosmological and astrophysical scales, but more realistic Lagrangians have to be taken into account than simple power-law ones. However, in our case, this model provides a singularity-free bouncing solution while avoiding the Dolgov-Kawasaki instability. This feature is significant as it addresses common issues in cosmological models, offering a stable and consistent framework for studying early universe dynamics and potentially contributing to our understanding of cosmic inflation and bounce scenarios. Our model's flexibility to encompass non-phantom, phantom, and $\Lambda$CDM cases provides a broad framework to test other models. However, specific parameter constraints could rule out models that deviate significantly from these values mathematically. However, one needs to test the above models with the new upcoming observational data for a better study. Furthermore, in our oscillatory bouncing case, we obtained a new \( f(\mathcal{R}) \) model that incorporates a hypergeometric function, which is relatively unexplored in the literature. However, in our case, this model provides a singularity-free bouncing solution while avoiding the Dolgov-Kawasaki instability. This feature is significant as it addresses common issues in cosmological models, offering a stable and consistent framework for studying early universe dynamics and potentially contributing to our understanding of cosmic inflation and bounce scenarios.

In this study, a notable advancement is the explicit analytical derivation of the $f(\mathcal{R})$ gravity model. The models accommodate early-time bouncing behavior and late-time cosmic acceleration within a unified framework. Such findings provide valuable insight into constructing models that depict the early universe and offer a pathway for understanding the cosmological mechanism driving the ongoing accelerated expansion of the universe.
It would be interesting to find further motivated reconstructed models in Palatini's theories and other aspects of quantum gravity phenomenology using the same approach in future works. We also aim to construct a viable non-singular bounce in $f(\mathcal{R})$ gravity using the above technique in an inflationary context.

\section*{Data Availability}
Data sharing not applicable to this article as no datasets were generated or analysed during the current study.

\section*{Acknowledgments}
GNG acknowledges University Grants Commission (UGC), New Delhi, India for awarding a Senior Research Fellowship (UGC-Ref. No.: 201610122060). PKS acknowledges Science and Engineering Research Board, Department of Science and Technology, Government of India for financial support to carry out Research project No.: CRG/2022/001847 and IUCAA, Pune, India for providing support through the visiting Associateship program. The work of KB is supported in part by the JSPS KAKENHI Grant Number 21K03547 and 23KF0008. We are very much grateful to the honourable referee and to the editor for the illuminating suggestions that have significantly improved our work in terms of research quality and presentation. 



\end{document}